 \documentclass[final,3p,times]{elsarticle}
\usepackage{amsfonts}
\usepackage{amsmath}
\usepackage{graphicx}

\usepackage{lineno}

\begin{document}
\begin{frontmatter} 
\title{The Complexity of Shelflisting}
\author{Yongjie Yang and Dinko Dimitrov \\
{\small Chair of Economic Theory, Saarland University, Germany}}

\begin{abstract}
Optimal shelflisting invites profit maximization to become sensitive to the
ways in which purchasing decisions are order-dependent. We study the
computational complexity of the corresponding product arrangement problem
when consumers are either rational maximizers, use a satisficing procedure,
or apply successive choice. The complexity results we report are shown to
crucially depend on the size of the top cycle in consumers' preferences over
products and on the direction in which alternatives on the shelf are
encountered.\medskip \newline
\textit{JEL} Classification Number: D00; D01; D03; D11; D20\newline
\end{abstract}
\end{frontmatter} 

\section{Introduction}

There are at least two main research directions in recent works in economics
devoted to the study of order or frame effects on consumers' behavior. The
first one adopts a choice theoretic approach and provides foundations for
encompassing non-standard behavior models (cf. Masatlioglu and Ok, 2005;
Rubinstein and Salant, 2006; Salant and Rubinstein, 2008; Bernheim and
Rangel, 2009), while the second direction incorporates frames as part of
players' strategy spaces and analyzes the structure of the corresponding
market equilibrium outcomes (cf. Eliaz and Spiegler, 2011; Spiegler, 2014).
A complementary viewpoint is provided by experimental studies in the
marketing literature (cf. Valenzuela and Raghubir, 2009; Valenzuela et al.,
2013) with the focus being on consumers' beliefs about the organization of
product displays and their impact on position-based consumers' preferences
over products as well as on retailers' actual shelflistings.

Our starting point in the present paper is that of a single shelf designer
who has to arrange a given number of products on a shelf in a way that
maximizes his profit. In doing so, he is facing a finite set of consumers
who select a single product from the shelf by using a pre-specified choice
rule. However, in sharp contrast to the cited works, we analyze the effects
of consumers' behavior on the optimal shelf listing from a computational
complexity perspective.\footnote{%
We will assume familiarity with basic concepts in computational complexity:
exponential time, polynomial time, polynomial-time reductions, and
NP-hardness.} For this, we set the shelf designer's product arrangement
problem (denoted by \textsc{PA}) and study its computational complexity when
consumers are either rational maximizers, follow a satisficing procedure for
choosing from lists, or apply successive choice when purchasing their
products (see Section 2 for the corresponding definitions).

We show that this decision problem is computationally easy when consumers
are rational (Theorem 1), while turning to be \textit{in general} more
difficult when they use a satisficing choice rule (Theorems 2 and 3) as then
to become hard for the case of successive choice (Theorems 4-7). The term
\textquotedblleft in general\textquotedblright\ stands as to indicate the
sensibility of our results with respect to the following features of the
decision problem. First, we allow a consumer to encounter the products on
the shelf either from left to right or from right to left (the product
arrangement problem when all consumers check the list in the same direction
(from left to right) is denoted by \textsc{SE-PA}). Second, in the case of
successive choice, allowing for non-transitive consumer preferences makes our results dependent on the size of the corresponding top
cycles, that is, on the number of consumers' top favorite products. This
dependence is summarized in Table 1.

Our focus on the three types of consumers' behavior is partially in line
with the corresponding findings in Salant (2011) with respect to their
\textit{state} complexity. Since the problem we study is from the viewpoint
of a shelf designer \textit{facing} various ways of product selection from
lists (and we focus on its \textit{computational} complexity), it is natural to
expect that the corresponding statements go in rather opposite directions.
For instance, while the state (or procedural) complexity of rational choice
is much higher than the one of satisficing choice (see Propositions 2 and 3
in Salant, 2011), we show that the \textsc{PA} problem in the latter case
is NP-hard, while being polynomial-time solvable in the former case. The
corresponding statement with respect to the successive choice rule also
applies. Notice additionally, that our results do also partially allow for a
comparison of the complexity of the \textsc{SE-PA} problem when order
effects are taken into account (that is, when consumers are satisficers or
use successive choice). More precisely, as $\log \frac{n}{n-1}$ is smaller
than $1$ for large values of $n$, $O(n^{2}m)$ is asymptotically greater than
$O(n(\log n+m))$. So, our algorithm to solve the \textsc{SE-PA} problem when
consumers are satisficers is asymptotically faster than the one to solve
this problem when there are at most three top favorite products for each
consumer and successive choice is applied.

\begin{table*}[tbp]
\begin{center}
\begin{tabular}{|c|c|c|c|c|c|}
\hline
Decision & Rational  & Satisficing &
\multicolumn{3}{c|}{Successive choice} \\ \cline{4-6}
problem& choice & choice & $t=1$ & $t=3$ & $t\geq 4$ \\ \hline
\textsc{PA} & $O(m)$ & NP-hard & $O(nm)$ & {NP-hard} & NP-hard \\ \hline
\textsc{SE-PA} & $O(m)$ & $O(n(\log {n}+m))$ & $O(nm)$ & $O(n^{2}m)$ &
NP-hard \\ \hline
\end{tabular}%
\end{center}
\caption{A summary of our results. In the entries corresponding to
successive choice, $t$ is the upper bound of the number of top favorite
products of each buyer. Here $n$ is the number of products and $m$ is the
number of buyers.}
\label{tab_summary_results}
\end{table*}

The rest of the paper is structured as follows. In Section 2 we provide the
basic definitions with respect to lists, consumer preferences and the choice
functions we consider. Section 3 is then devoted to the problem formulation
and the case when consumers are rational maximizers, while Section 4
contains our results for consumers following a satisficing procedure for
product selection. Section 5 contains then the ways in which the difficulty
of a shelf designer's task depends on the number of consumers' top favorite
products as well as on the direction in which they encounter the listed
alternatives.

\section{Lists, preferences, and choices}

Our setup consists of the following basic ingredients.\medskip \newline
\textbf{Products and lists}\medskip

Let $P=\{p_{1},p_{2},...,p_{n}\}$ be a set of $n$ products. A \textit{list}
is a permutation of $P$. We denote by $\mathcal{L}(P)$ the set of all $n!$
lists over $P$. For convenience, for each list $L=(p_{\beta (1)},p_{\beta
(2)},\ldots ,p_{\beta (n)})\in \mathcal{L}(P)$ we refer to $p_{\beta (1)}$
as the \textit{leftmost product} and to $p_{\beta (n)}$ as the \textit{%
rightmost product}. Moreover, for each $p_{\beta (i)}$ with $i\in \{1,\ldots
,n\}$, we refer to $i$ as the \textit{position} of the product $p_{\beta
(i)} $ in the list $L$. Denoting by $B=\{b_{1},\ldots ,b_{m}\}$ the set of $m
$ \textit{buyers} (consumers), we assume that each consumer is purchasing
the product delivered by her pre-specified choice rule from lists.\medskip\
\newline
\textbf{Preferences}\medskip

A \textit{tournament preference} is defined as a complete and asymmetric
binary relation $\succ $ over $P$. In particular, $p_{i}\succ p_{j}$
signifies that $p_{i}$ is \textit{preferred} to $p_{j}$. We say that $p_{i}$
is weakly preferred over $p_{j}$ if either $p_{i}$ is preferred to $p_{j}$
or $p_{i}=p_{j}$. The \textit{top circle} of $\succ $, denoted by $TC(\succ
) $, is the unique subset $P^{\prime }\subseteq P$ of minimum cardinality
such that every product in $P^{\prime }$ is preferred to every product not
in $P^{\prime }$. For a buyer with tournament preference $\succ $, products
in $TC(\succ )$ are called her \textit{top favorite} products. The set of
all tournament preferences is denoted by $\mathcal{T}(P)$. If we require a
tournament preference to be transitive, then we have a \textit{linear
preference}. Clearly, for each linear preference $\succ $, $\left\vert
TC(\succ )\right\vert =1$. The unique element in $TC(\succ )$ in this case
is referred to as the \textit{top product} according to $\succ $. Notice
that a linear preference can be seen as a permutation over $P$. Precisely,
the position of a product $p$ according to $\succ $ is $\left\vert p^{\prime
}\in P:p^{\prime }\succ p\right\vert +1$. Hence, $\mathcal{L}(P)$ is the set
of all linear preferences.\medskip \newline
\textbf{Choice functions}\medskip

We assume that each consumer purchases the product determined by the outcome
of her corresponding choice function. Choice functions from lists were
formally introduced in Rubinstein and Salant (2006) (see also Simon, 1955
and Salant, 2003) in order to describe choice behavior potentially affected
by the order of the products in a list. We first describe below the three
families of such functions we consider as then to provide a more general
description.

\begin{description}
\item[\textit{{Rational choice}} $(f^{RC})$.] Each buyer has a linear
preference $\succ $ $\in \mathcal{L}(P)$ and chooses her most preferred
product according to $\succ $, that is, the one in $TC(\succ )$.

\item[\textit{{Satisficing choice}} $(f^{SAT})$.] Each buyer $b$ has a
tournament preference $\succ _{b}\in \mathcal{T}(P)$ and a threshold product
$p(b)\in P$. The buyer chooses the first encountered product from a list
that is weakly preferred to her threshold product.
\end{description}

\begin{description}
\item[\textit{{Successive choice }}$\left( f^{SC}\right) $.] Each buyer has
a tournament preference $\succ $ $\in \mathcal{T}(P)$ and chooses her
product as follows. The buyer first stores her first encountered product in
a register. Then, she goes through the products further and compares the
currently encountered product $p_{j}$ with the one ($p_{i}$) in the
register. If $p_{j}\succ p_{i}$, then she replaces $p_{i}$ by $p_{j}$ in the
register and goes forward; otherwise, she goes forward without changing the
product in the register. After the buyer encounters all products, she
purchases the product in the register.
\end{description}

Clearly, a rational choice function is independent of any order effects in a
list, while the other two choice functions are sensitive with respect to
such order effects. Notice for instance that a consumer $b$ using a
satisficing procedure is more likely to select $p(b)$ when one moves $p(b)$
toward the beginning of a list and thus, primacy effect is displayed. The
magnitude of such a primacy effect crucially depends on the fact whether a
consumer starts encountering the products in a list from the left side to
the right side, or she checks the products from the right side to the left
side. In order to pay attention to both ways for encountering the products
on a shelf, we will use the following description of choice functions from
lists discussed above.

For each $X\in \{RC,SC\}$, we define $f^{X}:\mathcal{L}(P)\times {\mathfrak{P%
}}\times \{1,0\}\rightarrow P$ to be a function assigning a single element $%
f^{X}(L,Pref,v)\in P$ to every list $L=(p_{1},...,p_{n})\in \mathcal{L}(P)$,
every preference ${{Pref}\in \mathfrak{P}}$ in the corresponding domain
(i.e., $\mathfrak{P}=\mathcal{L}(P)$ for $X=RC$ and $\mathfrak{P}=\mathcal{T}%
(P)$ for $X=SC$), and every $v\in \{1,0\}$. The third component $v\in
\{1,0\} $ indicates from which side the corresponding buyer begins to go
through the products in the list $L$. In particular, \textquotedblleft $v=1$%
\textquotedblright\ means that the buyer goes through the products from the
left side to the right side, and \textquotedblleft $v=0$\textquotedblright\
means the buyer goes through the products from the right side to the left
side. We call buyers in the former case \textit{left-biased}, and buyers in
the latter case \textit{right-biased}. The choice function $f^{SAT}:\mathcal{%
L}(P)\times {\mathcal{T}(P)}\times \{1,0\}\times P\rightarrow P$ assigns a
single element $f^{SAT}(L,\succ ,v,p)\in P$ to every list $L$, every
tournament preference ${\succ }\in \mathcal{T}(P)$, every $v\in \{1,0\}$ and
every $p\in P$. Here, the fourth component indicates the threshold product
of a buyer.

\section{Problem formulation and easiest shelflistings}

As already indicated in the Introduction, the shelf designer's problem is to
determine a product arrangement by taking into account consumers' choice
rules. More precisely, we will consider the complexity of the following
problem for each $f^{X}$ with $X\in \{RC,SAT,SC\}$ by letting $\mathfrak{P}=%
\mathfrak{\mathcal{L}}(P)$ for $X=RC$ and $\mathfrak{P}=\mathcal{T}(P)$ for $%
X\in \{SAT,SC\}$.

\begin{description}
\item[{\textbf{Product Arrangement}} \textsc{({PA-$f^{X}$})}]

\item[Input] A set $P=\{p_{1},p_{2},\ldots ,p_{n}\}$ of $n$ products each of
which has infinite supplies, a profit function $\mu :P\rightarrow \mathbb{R}%
^{+}$, a set $B=\{b_{1},b_{2},\ldots ,b_{m}\}$ of $m$ buyers where each $%
b_{i}$ is associated with a preference $Pref_{b_{i}}\in \mathfrak{P}$, an
entering function $\omega :B\rightarrow \{0,1\}$, and a real number $R$. For
$f^{SAT}$, each buyer $b_{i}\in B$ is associated with a threshold product $%
p(b_{i})$.

\item[Question] Is there a list $L\in \mathcal{L}(P)$ such that%
\begin{equation*}
\sum_{b_{i}\in B}\mu \left( f^{X}(L,Pref_{b_{i}},\omega (b_{i}))\right) \geq
R
\end{equation*}%
for $X\in \{RC,SC\}$, and%
\begin{equation*}
\sum_{b_{i}\in B}\mu \left( f^{X}(L,Pref_{b_{i}},\omega
(b_{i}),p(b_{i}))\right) \geq R
\end{equation*}%
for $X=SAT$?
\end{description}

In the above definition, the value of $\omega (b_{i})$ indicates whether $%
b_{i}$ is left-biased ($\omega (b_{i})=1$) or right-biased ($\omega
(b_{i})=0 $). For each $p\in P$, $\mu (p)$ is the profit associated with the
product $p $ when sold.

We will also consider a special case of the above problem where $\omega
(b_{i})=1$ for all $b_{i}\in B$. We denote this problem as \textsc{%
Single-Enter Product Arrangement} (\textsc{{SE-PA}-$f^{X}$}). For
simplicity, in this case, we drop $\omega (b_{i})$ in the above definition.

Notice that buyers using the rational choice function choose their products
regardless of how the products are arranged in a list. This directly leads
to the following theorem.\medskip \newline
\textbf{Theorem 1} \textit{SE-PA-}$f^{RC}$\textit{\ and PA-}$f^{RC}$\textit{%
\ are solvable in }$O(m)$\textit{-time.\medskip }\newline
\textbf{Proof.} Since each buyer chooses her most preferred product
regardless of how the products are arranged in the list and from which
direction she encounters the products, to solve the problems stated in the
theorem we need only to sum up the profits of the most preferred products of
the buyers and compare the sum with $R$. If the sum is greater than or equal
to $R$, return \textquotedblleft YES\textquotedblright ; otherwise, return
\textquotedblleft NO\textquotedblright . Since we have in total $m$ buyers
and it takes $O(1)$ time to calculate $TC(\succ )$ for each linear
preference $\succ $, the algorithm terminates in $O(m)$ time. \rule{2mm}{2mm}

\section{Biased consumers and satisficing choice}

Let us now turn to the situation where consumers use a satisficing procedure
for selecting their products from the shelf. As it turns out, the decision
problem is polynomial-time solvable, provided that all consumers are
encountering the alternatives in the same direction (Theorem 2), while
becoming NP-hard when both left-biased and right-biased consumers are allowed (Theorem 3).\medskip
\newline
\textbf{Theorem 2 }\textit{SE-PA-}$f^{SAT}$\textit{\ is solvable in }$%
O(n(\log n+m))$\textit{\ time.\medskip }\newline
\textbf{Proof. }We prove the theorem by developing a polynomial-time
algorithm of corresponding running time for the problem stated in the
theorem. In what follows, for each buyer $b\in B$, let $\succ _{b}$ and $%
p(b) $ be the tournament preference and the threshold product of $b$,
respectively.

The algorithm is quite trivial: sort the products according to the profits,
from the highest to the lowest, and determine if this results in a solution.
Formally, let $L$ be a list $\left( p_{\pi (1)},\ldots ,p_{\pi (n)}\right) $
such that $\mu (p_{\pi (x)})\geq \mu (p_{\pi (y)})$ for every $1\leq x<y\leq
n$. Then, if $\sum_{b\in B}\mu \left( f^{SAT}(L,\succ _{b},p(b))\right)
\geq R$, return \textquotedblleft YES\textquotedblright ; otherwise return
\textquotedblleft NO\textquotedblright . Such a list $L$ can be constructed
in $O(n\log n)$ time by the Merge sort algorithm (cf. Katajainen and Tr\"{a}%
ff, 1997). In addition, it takes $O(nm)$ time for all buyers to go through $L
$ and choose their products. Calculating the sum of the profits of the
chosen products takes $O(m)$ time. In total, the above algorithm takes $%
O(n(\log n+m))$ time. It remains to prove the correctness of the algorithm.
For this, the following claim will be useful.\medskip \newline
\textbf{Claim} Let $L_{1}=\left( p_{\alpha (1)},\ldots ,p_{\alpha
(n)}\right) $ be a list, where $\alpha $ is a permutation of $\left\{
1,2,\ldots ,n\right\} $, and let, for $x\in \left\{ 1,2,\ldots ,n-1\right\} $%
, $p_{\alpha (x)}$ and $p_{\alpha (x-1)}$ be two consecutive products with $%
\mu (p_{\alpha (x)})>\mu (p_{\alpha (x-1)})$. Let $L_{2}$ be the list
obtained from $L_{1}$ by swapping $p_{\alpha (x)}$ and $p_{\alpha (x-1)}$.
Then, it holds that%
\begin{equation*}
\sum_{b\in B}\mu \left( f^{SAT}(L_{2},\succ _{b},p(b))\right) \geq
\sum_{b\in B}\mu \left( f^{SAT}(L_{1},\succ _{b},p(b))\right) .
\end{equation*}%
\newline
\textbf{Proof of the Claim.} Let $b\in B$ be any arbitrary buyer. If there
is an $x^{\prime }$ with $1\leq x^{\prime }<x-1$ such that $p_{\alpha
(x^{\prime })}\succ p(b)$ or for every $y$ with $1\leq y\leq x$ it holds
that $p(b)\succ p_{\alpha (y)}$, then $\mu \left( f^{SAT}(L_{1},\succ
_{b},p(b))\right) =\mu \left( f^{SAT}(L_{2},\succ _{b},p(b))\right) $.
Otherwise, either $p_{\alpha (x-1)}$ or $p_{\alpha (x)}$ is the first
encountered product that is preferred to $p(b)$, or equivalently, $%
f^{SAT}(L_{1},\succ _{b},p(b))$, $f^{SAT}(L_{2},\succ _{b},p(b))\in \left\{
p_{\alpha (x-1)},p_{\alpha (x)}\right\} $. If $p_{\alpha (x)}\succ p(b)$,
then $f^{SAT}(L_{2},\succ _{b},p(b))=p_{\alpha (x)}$ and $%
f^{SAT}(L_{1},\succ _{b},p(b))\in \left\{ p_{\alpha (x-1)},p_{\alpha
(x)}\right\} $; otherwise, $f^{SAT}(L_{2},\succ
_{b},p(b))=f^{SAT}(L_{1},\succ _{b},p(b))=p_{\alpha (x-1)}$. Therefore, we
have in both cases that $\mu \left( f^{SAT}(L_{2},\succ _{b},p(b))\right)
\geq \mu \left( f^{SAT}(L_{1},\succ _{b},p(b))\right) $. It then directly
follows that \[\sum_{b\in B}\mu \left( f^{SAT}(L_{2},\succ
_{b},p(b))\right) \geq \sum_{b\in B}\mu \left( f^{SAT}(L_{1},\succ
_{b},p(b))\right).\] This completes the proof of the claim.\medskip

Due to the above claim, if we have a solution $L$, we can obtain another
solution by swapping two consecutive products in $L$ as indicated in the
proof of the claim. By exhaustively performing this swap operation, we can
arrive at a list $\left( p_{\pi (1)},\ldots ,p_{\pi (n)}\right) $ such that $%
\mu (p_{\pi (x)})\geq \mu (p_{\pi (y)})$ for every $1\leq x<y\leq n$. The
correctness of the algorithm then follows. \rule{2mm}{2mm}\medskip

Let us now turn to the study of the PA-$f^{SAT}$ problem for the case where
there are both left-biased buyers and right-biased buyers. In a sharp
contrast to the results presented so far, we show that PA-$f^{SAT}$ is
NP-hard \textit{even when} buyers' preferences are restricted to be linear.
Our proof is by reduction from the following problem.

\begin{description}
\item[Restricted Betweenness]

\item[Input] Two disjoint sets $U=\{u_{1},...,u_{x}\}$, $V=\{v_{1},...,v_{y}%
\}$, an additional element $w\not\in U\cup V$, a collection $\mathcal{C}$ of
3-tuples $(u_{i},v_{j},u_{k})$ such that $1\leq i\neq k\leq x$ and $1\leq
j\leq y$, and a collection $\mathcal{D}$ of 3-tuples $(v_{i},w,u_{j})$ with $%
1\leq i\neq j\leq x$.

\item[Question] Is there a linear order over $U\cup V\cup \{w\}$ such that
for every $(a,b,c)\in \mathcal{C}\cup \mathcal{D}$ the element $b$ lies
between $a$ and $c$?
\end{description}

As the above problem has been shown to be NP-hard (cf. Opatrny, 1979), we
cannot expect to have an efficient algorithm to solve the PA-$f^{SAT}$
problem exactly, unless P=NP which is commonly believed to be
unlikely.\medskip\ \newline
\textbf{Theorem 3} \textit{PA-}$f^{SAT}$\textit{\ is NP-hard even when all
consumers have linear preferences.\medskip }\newline
\textbf{Proof.} Let $I=\left( U,V,w,\mathcal{C},\mathcal{D}\right) $ be an
instance of the \textsc{Restricted Betweenness} problem. We create first an
instance $I^{\prime }$ for the problem stated in the theorem.

The products are as follows. For each $u\in U$, we create a product $p(u)$
such that $\mu (p(u))=2$. For each $v\in V$, we create a product $p(v)$ such
that $\mu (p(v))=1$. In addition, we create for $w$ a product $p(w)$ such
that $\mu (p(w))=0$. Hence, the set of products created is $P=\left\{
p(a):a\in U\cup V\right\} \cup \left\{ p(w)\right\} $.

The buyers are as follows. For each 3-tuple $s=(a,b,c)\in \mathcal{C}\cup
\mathcal{D}$, we have one left-biased consumer $b_{s}^{\ell }$ with
threshold product $p(b_{s}^{\ell })$ and one right-biased consumer $%
b_{s}^{r} $ with threshold product $p(b_{s}^{r})$. Moreover, each of these
two buyers prefers each of the products in the set $\left\{
p(a),p(b),p(c)\right\} $ over her threshold product, while all remaining
products are ordered below it.

According to the above construction, the buyer $b_{s}^{\ell }$ (resp., $%
b_{s}^{r}$) corresponding to $s=(a,b,c)\in \mathcal{C}\cup \mathcal{D}$
chooses her first encountered product in $\left\{ p(a),p(b),p(c)\right\} $.
More precisely, given a list $L$, the buyer \ $b_{s}^{\ell }$ chooses the
leftmost product among $\left\{ p(a),p(b),p(c)\right\} $ and the buyer $%
b_{s}^{r}$ chooses the rightmost product among $\left\{
p(a),p(b),p(c)\right\} $ from the list $L$. Finally, we set the threshold
bound $R=4\left\vert \mathcal{C}\right\vert +3\left\vert \mathcal{D}%
\right\vert $.

The construction clearly takes polynomial time. It remains to prove the
correctness of the reduction.

($\Rightarrow :$) Let $L^{\prime }$ be a linear order over $U\cup V\cup
\left\{ w\right\} $ such that for every $(a,b,c)\in \mathcal{C}\cup \mathcal{%
D}$ the element $b$ is between $a$ and $c$. Let $L$ be the list obtained
from $L^{\prime }$ by replacing every element $a$ of $L^{\prime }$ by the
corresponding product $p(a)$. Due to the above discussion, for each $%
s=(u_{i},v_{j},u_{k})\in \mathcal{C}$, the two corresponding buyers $%
b_{s}^{\ell }$ and $b_{s}^{r}$ choose exactly the products $p(u_{i})$ and $%
p(u_{k})$, one for each buyer. As $\mu (p(u))=2$ for every $u\in U$, the
total profit of the products chosen by all buyers corresponding to 3-tuples
in $\mathcal{C}$ is exactly $4\left\vert \mathcal{C}\right\vert $.
Analogously, for each $s=(v_{i},w,u_{j})\in \mathcal{D}$, the two
corresponding buyers $b_{s}^{\ell }$ and $b_{s}^{r}$ choose exactly the
products $p(v_{i})$ and $p(u_{j})$ with $u_{j}\in U$ and $v_{i}\in V$. As $%
\mu (p(v))=1$ for every $v\in V$, the total profit of the products chosen by
all buyers corresponding to 3-tuples in $\mathcal{D}$ is exactly $%
3\left\vert \mathcal{D}\right\vert $. Hence, the total profit of the
products chosen by all buyers is $R=4\left\vert \mathcal{C}\right\vert
+3\left\vert \mathcal{D}\right\vert $.

($:\Leftarrow $) Observe that to achieve the total profit $R$, every pair of
buyers corresponding to a 3-tuple in $\mathcal{C}$ must choose two products
whose total profit is at least $4$, and every pair of buyers corresponding
to a 3-tuple in $\mathcal{D}$ must choose two products whose total profit is
at least $3$. Due to above discussion, this happens only if there is a list $%
L$ such that for every $s=(a,b,c)\in \mathcal{C}\cup \mathcal{D}$ the
product $p(b)$ is between the products $p(a)$ and $p(c)$. This implies that
the linear order obtained from $L$ by replacing every $p(a)$ with the
corresponding element $a\in U\cup V\cup \left\{ w\right\} $ is a solution of
$I$. \rule{2mm}{2mm}

\section{Top cycles and successive choice}

Recall that consumers' preferences in the definition of the successive
choice from lists were not necessarily restricted to be transitive, that is,
they may contain cycles. Clearly, since consumers go throughout the products
in the entire list, the existence (and, as it turns out, the size) of top
cycles matters. Lemma 1 stated below relates the consumer's choice from a
list when her corresponding top cycle is of size 3. More precisely, it is
always the last encountered product from the top favorite products that is
selected. We use then Lemma 1 and the construction in the proof of Theorem 3
in order to show that the product arrangement problem is NP-hard even when
each buyer has 3 most favorite products (Theorem 4).\medskip \newline
\textbf{Lemma 1} \textit{Let }$L=(p_{1},p_{2},\ldots ,p_{n})$\textit{\ be a
list over }$P$\textit{\ and }$b\in B$\textit{\ a buyer with preference }$%
\succ $\textit{. If }$TC(\succ )=\{p_{i},p_{j},p_{k}\}$\textit{\ with }$%
i<j<k $\textit{, then }$f^{SC}(L,\succ ,1)=p_{k}$\textit{\ and }$%
f^{SC}(L,\succ ,0)=p_{i}$\textit{.\medskip \newline
}\textbf{Proof.} Consider first the situation corresponding to $%
f^{SC}(L,\succ ,1)=p_{k}$. Clearly, exactly one product from the set $%
\left\{ p_{i},p_{j}\right\} $ is preferred to $p_{k}$. For each $x$ with $%
1\leq x\leq n$, let $register(x)$ be the product in the register immediately
after $p_{x}$ has been encountered. Hence, $register(1)=p_{1}$. We
distinguish between the following two cases corresponding to the two
possible preference cycles over the product set $\{p_{i},p_{j},p_{k}\}$%
.\medskip \newline
\textit{Case A} ($p_{i}\succ p_{k}$). Clearly, it must be then that $%
p_{k}\succ p_{j}$ and $p_{j}\succ p_{i}$ holds. It is easy to check that $%
register(j-1)=p_{i}$, $register(j)=register(k-1)=p_{j}$, and $%
register(k)=register(n)=p_{k}$. This implies that $f^{SC}(L,\succ ,1)=p_{k}$%
.\medskip \newline
\textit{Case B} ($p_{k}\succ p_{i}$). We have now that $p_{i}\succ p_{j}$
and $p_{j}\succ p_{k}$ should hold. It follows that $%
register(j-1)=register(j)=register(k-1)=p_{i}$ and $%
register(k)=register(n)=p_{k}$. This again implies that $f^{SC}(L,\succ
,1)=p_{k}$.\medskip

The proof for the situation corresponding to $f^{SC}(L,\succ ,0)=p_{i}$ can
be obtained from the above proof by swapping all occurrences of $p_{i}$ and $%
p_{k}$, and replacing all occurrences of $1$ by $0$. \rule{2mm}{2mm}\medskip
\newline
\textbf{Theorem 4} \textit{PA-}$f^{SC}$\textit{\ is NP-hard, even when each
buyer has }$3$\textit{\ top favorite products.\medskip }\newline
\textbf{Proof.} We can utilize the construction used in the proof of Theorem
3 with the following slight difference when creating the buyers. For each
3-tuple $s=(a,b,c)\in \mathcal{C}\cup \mathcal{D}$, we have one left-biased
consumer $b_{s}^{\ell }$ (i.e., $\omega (b_{s}^{\ell })=1$) and one
right-biased consumer $b_{s}^{r}$ (i.e., $\omega (b_{s}^{r})=0$) whose top
favorite products are $p(a)$, $p(b)$, and $p(c)$. Clearly, for each of these
buyers, we have either $p(a)\succ p(b)\succ p(c)\succ p(a)$ or $p(b)\succ
p(a)\succ p(c)\succ p(b)$ with each of these three products being preferred
to any of the remaining products. The application of Lemma 1 gives us then
the same consumer choices as the ones in the proof of Theorem 3. \rule%
{2mm}{2mm}\medskip

Our next lemma states that the outcome of the successive choice rule is
always a top favorite product for the corresponding consumer (the proof is
straightforward). It then follows via Lemma 2 that if each buyer has only
one favorite product, then the PA-$f^{SC}$ problem becomes polynomial-time
solvable (Theorem 5).\medskip \newline
\textbf{Lemma 2 }\textit{Let }$L=(p_{1},p_{2},...,p_{n})$\textit{\ be a list
over }$P$\textit{\ and }$b\in B$\textit{\ be a buyer with preference }$\succ
$\textit{. Then, }$f^{SC}(L,\succ ,\omega (b))\in TC(\succ )$\textit{%
.\medskip }\newline
\textbf{Theorem 5} \textit{PA-}$f^{SC}$\textit{\ is solvable in }$O(nm)$%
\textit{\ time if each buyer has only one top favorite product.\medskip
\newline
}\textbf{Proof. }Due to Lemma 2, each buyer chooses the product in the top
circle of her preference no matter how the products are arranged in the
list. Hence, we can directly adopt the algorithm for PA-$f^{RC}$. However,
since it takes $O(n)$ time to calculate the top circle in this case, the
algorithm has running time $O(nm)$.\textbf{\ \rule{2mm}{2mm}\medskip }

Consider now SE-PA-$f^{SC}$, the special case of PA-$f^{SC}$ where there
are only left-biased consumers. We first show that if each buyer has at most
3 top favorite products, then the problem is polynomial-time
solvable.\medskip \newline
\textbf{Theorem 6} \textit{SE-PA-}$f^{SC}$\textit{\ is solvable in }$%
O(n^{2}m)$\textit{\ time if }$\left\vert TC(\succ _{b})\right\vert \leq 3$%
\textit{\ for every buyer }$b\in B$\textit{, where }$\succ _{b}$\textit{\ is
the preference of }$b$\textit{.\medskip \newline
}\textbf{Proof. }For $P^{\prime }\subseteq P$, let $\mu _{\max }(P^{\prime
})=\max \left\{ \mu (p):p\in P^{\prime }\right\} $. Consider the following
algorithm: if $\sum _{b\in B}\mu _{\max }(TC(\succ _{b}))\geq R$, return
\textquotedblleft YES\textquotedblright ; otherwise return \textquotedblleft
NO\textquotedblright . We show that the algorithm correctly solves the SE-PA-%
$f^{SC}$ problem.

Let $A$ be the set of buyers $b$ such that $\left\vert TC(\succ
_{b})\right\vert =1$, and $C$ the set of buyers $b$ such that $\left\vert
TC(\succ _{b})\right\vert =3$. Due to Lemma 2, every buyer in $A$ chooses
the product in $TC(\succ _{b})$ for every list $L$ over $P$. So, the total
profits from the products chosen by the buyers in $A$ is $\sum_{b\in
A}\mu (TC(\succ _{b}))=\sum_{b\in A}\mu _{\max }(TC(\succ _{b}))$. Let $%
L=\left( p_{\pi (1)},\ldots ,p_{\pi (n)}\right) $ be a list such that $\mu
(p_{\pi (x)})\leq \mu (p_{\pi (y)})$ for every $x$ and $y$ such that $1\leq
x<y\leq n$. Consider now a buyer $b\in C$. Without loss of generality,
assume that $TC(\succ _{b})=\left\{ p_{\pi (x)},p_{\pi (y)},p_{\pi
(z)}\right\} $ with $x<y<z$. Due to Lemma 1, $f^{SC}(L,\succ _{b})=p_{\pi
(z)}$ with $\mu \left( f^{SC}(L,\succ _{b})\right) =\mu _{\max }(TC(\succ
_{b}))$ following from the definition of $L$. In summary, the list $L$
results in the highest total profit that can be achieved, since with respect
to $L$ every buyer $b$ chooses a product with a highest profit among all
products that are possible for the buyer to choose. The correctness of the
algorithm follows.

It remains to show the running time of the algorithm. Calculating $TC(\succ
_{b})$ for every $b\in B$ can be done in $O(n^{2})$ time (cf. Tarjan, 1972
and Sharir, 1981). Since $TC(\succ _{b})$ is of size at most $3$, we can
calculate each $\mu _{\max }(TC(\succ _{b}))$ in $O(1)$ time. Summing up
them takes $O(m)$ time. Since we have $m$ buyers, the whole running time of
the algorithm is bounded by $O(n^{2}m)+O(m)=O(n^{2}m)$. \textbf{\rule%
{2mm}{2mm}\medskip }

\begin{figure}[h!]
\begin{center}
\includegraphics[width=0.95\textwidth]{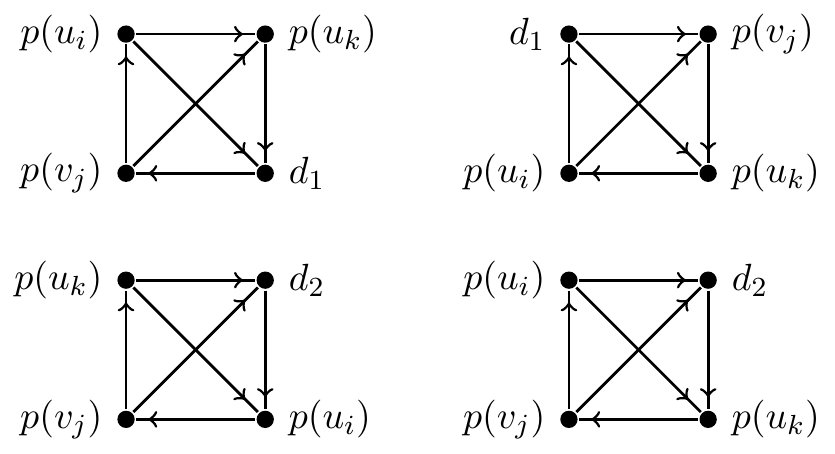}
\end{center}
\caption{This figure shows the top favorite products and the preferences of the four buyers created for a $(u_i,v_j,u_k)\in \mathcal{C}$ in the NP-hardness of the SE-PA-$f^{SC}$ problem in Theorem~7. An arc from a product to another product means that the former one is preferred to the latter one.}
\label{fig_NP_top_circle_4}
\end{figure}

We prove now that if we allow each buyer to have one more top favorite
product, the problem becomes NP-hard. The proof utilizes Figures~\ref{fig_NP_top_circle_4}-\ref{fig_NP_top_circle_4_b} and
Tables~\ref{tab_NP_hard_aone}-\ref{tab_NP_hard_two} which can be found in the Appendix.\medskip\ \newline
\textbf{Theorem 7} \textit{SE-PA-}$f^{SC}$\textit{\ is NP-hard even if each
buyer has at most }$4$\textit{\ top favorite products.\medskip }\newline
\textbf{Proof.} We prove the theorem by a reduction from the \textsc{%
Restricted Betweenness} problem. Let $I=\left( U,V,w,\mathcal{C},\mathcal{D}%
\right) $ be an instance of the \textsc{Restricted Betweenness} problem. We
create first an instance $I^{\prime }$ for the problem stated in the theorem.

The products are as follows. For each $u\in U$, we create a product $p(u)$
such that $\mu (p(u))=31$. For each $v\in V$, we create a product $p(v)$
such that $\mu (p(v))=32$. In addition, we create for $w$ a product $p(w)$
such that $\mu (p(w))=33$. Finally, we create two dummy products $d_{1}$ and
$d_{2}$ such that $\mu (d_{1})=1$ and $\mu (d_{2})=35$.

The buyers are as follows. For each 3-tuple $s=(u_{i},v_{j},u_{k})\in
\mathcal{C}$, we create $4$ buyers whose top favorite products and
preferences are as shown in Figure~\ref{fig_NP_top_circle_4}. Moreover, for each $%
s=(v_{i},w,u_{j})\in \mathcal{D}$, we create $14$ buyers whose top favorite
products and preferences are as shown in Figure~\ref{fig_NP_top_circle_4_b}. In total, we have $%
4\left\vert \mathcal{C}\right\vert +14\left\vert \mathcal{D}\right\vert $
buyers.

Finally, we set $R=129\left\vert \mathcal{C}\right\vert +464\left\vert
\mathcal{D}\right\vert $.

The construction clearly takes polynomial time. It remains to prove the
correctness of the reduction.

($\Rightarrow :$) Suppose that $I$ is a YES-instance. Let $L^{\prime }$ be a
solution of $I$, and $L$ be the list obtained from $L^{\prime }$ by first
replacing every $a\in L^{\prime }$ by $p(a)$ and then installing $d_{1}$ and
$d_{2}$ as $L$'s leftmost and rightmost products, respectively. Consider now
the profit from the products selected by the four buyers created for a $%
(u_{i},v_{j},u_{k})\in \mathcal{C}$. According to the constructions and
Tables~\ref{tab_NP_hard_aone}~and~\ref{tab_NP_hard_atwo}, the four buyers choose $p(v_{j})$, $p(u_{k})$, $p(u_{k})$, $d_{2}$,
respectively, if $u_{i}$ $L^{\prime }$ $v_{j}$ $L^{\prime }$ $u_{k}$, and
choose $p(v_{j})$, $p(u_{i})$, $d_{2}$, $p(u_{i})$, respectively, if $u_{k}$
$L^{\prime }$ $v_{j}$ $L^{\prime }$ $u_{i}$. Here $a$ $L^{\prime }$ $b$
means that $a$ is on the left side of $b$ in $L^{\prime }$. In both cases,
the total profit from these products is $32+31+31+35=32+31+35+31=129$.
Therefore, the total profit from the products chosen by buyers corresponding
to 3-tuples in $\mathcal{C}$ with respect to $L$ is $129\left\vert \mathcal{C%
}\right\vert $. Consider now a $\left( v_{i},w,u_{j}\right) \in \mathcal{D}$%
. According to the construction and Tables~\ref{tab_NP_hard_one} and~\ref{tab_NP_hard_two}, if $v_{i}$ $L^{\prime }$ $w$ $%
L^{\prime }$ $u_{j}$ then the $14$ buyers corresponding to $\left(
v_{i},w,u_{j}\right) $ choose the following products (number of buyers:
product):%
\begin{equation*}
\begin{array}{ll}
\mathbf{1:} & p(w); \\
\mathbf{7:} & d_{2}; \\
\mathbf{5+1=6:} & p(u_{j}).%
\end{array}%
\end{equation*}%
On the other side, if $u_{j}$ $L^{\prime }$ $w$ $L^{\prime }$ $v_{i}$ then
the $14$ buyers corresponding to $\left( v_{i},w,u_{j}\right) $ choose the
following products (number of buyers: product):%
\begin{equation*}
\begin{array}{ll}
\mathbf{1:} & p(w); \\
\mathbf{1+7=8:} & p(v_{i}); \\
\mathbf{5:} & d_{2}.%
\end{array}%
\end{equation*}%
Notice then that the total profit from these products is $464$ and thus, the
total profit from the choices of the buyers corresponding to 3-tuples in $%
\mathcal{D}$ is $464\left\vert \mathcal{D}\right\vert $. Hence, the total
profit from the products chosen by all buyers is exactly equal to $R$. We
conclude then that $I^{\prime }$ is a YES-instance.

($:\Leftarrow $) Suppose that $I^{\prime }$ is a YES-instance. Let $L$ be a
solution of $I^{\prime }$, and $L^{\prime }$ be obtained from $L$ by first
removing $d_{1}$ and $d_{2}$, and then replacing each product $p(a)$ with
its corresponding element in $U\cup V\cup \left\{ w\right\} $. Observe that
according to the construction, the maximum total profit from the choices by
the four buyers corresponding to a $\left( u_{i},v_{j},u_{k}\right) \in
\mathcal{C}$ is $129$. Moreover, this happens if and only if the products in
the top circles of the corresponding buyers' preferences fall into one of
the following cases (see Tables~\ref{tab_NP_hard_aone} and~\ref{tab_NP_hard_atwo} for further details):

\begin{itemize}
\item $p(u_{k})$ $L$ $d_{1}$ $L$ $p(v_{j})$ $L$ $p(u_{i})$ $L$ $d_{2};$

\item $d_{1}$ $L$ $p(u_{k})$ $L$ $p(v_{j})$ $L$ $p(u_{i})$ $L$ $d_{2};$

\item $d_{1}$ $L$ $p(u_{i})$ $L$ $p(v_{j})$ $L$ $p(u_{k})$ $L$ $d_{2};$

\item $p(u_{i})$ $L$ $d_{1}$ $L$ $p(v_{j})$ $L$ $p(u_{k})$ $L$ $d_{2}.$
\end{itemize}

The key point here is that $d_{2}$ has to be chosen at least once, since
otherwise the total profit from the products chosen by the four buyers
corresponding to a 3-tuple in $\mathcal{C}$ can be at most $32\times 4=128$.
Moreover, $d_{1}$ cannot be chosen by anyone, since otherwise the total
profit from the products chosen by the four buyers corresponding to a
3-tuple in $\mathcal{C}$ can be at most $35\times 3+1=106$. So, according to
Tables~\ref{tab_NP_hard_aone} and~\ref{tab_NP_hard_atwo}, only the above four cases and $\left( \left\{
p(u_{k}),p(v_{j})\right\} ,d_{1},p(u_{i}),d_{2}\right) $ satisfy these
conditions. However, given $\left( \left\{ p(u_{k}),p(v_{j})\right\}
,d_{1},p(u_{i}),d_{2}\right) $, the total profit from the products chosen by
the four buyers corresponding to $\left( u_{i},v_{j},u_{k}\right) $ is $\mu
(u_{i})+\mu (u_{k})+\mu (d_{2})+\mu (u_{i})=31+31+35+31=128<129$. A further
tedious check shows that all of the above cases lead to the same total
profit of $129$.

Notice further that the maximal total profit from the choices of the $14$
buyers corresponding to a $\left( v_{i},w,u_{j}\right) \in \mathcal{D}$ is $%
464$. Moreover, a tedious check shows that this happens if and only if the
products in the top circles of the corresponding buyers' preferences fall
into one of the following cases (see Tables~\ref{tab_NP_hard_one} and~\ref{tab_NP_hard_two} for further details):

\begin{itemize}
\item $p(u_{j})$ $L$ $d_{1}$ $L$ $p(w)$ $L$ $p(v_{i})$ $L$ $d_{2};$

\item $d_{1}$ $L$ $p(u_{j})$ $L$ $p(w)$ $L$ $p(v_{i})$ $L$ $d_{2};$

\item $d_{1}$ $L$ $p(v_{i})$ $L$ $p(w)$ $L$ $p(u_{j})$ $L$ $d_{2};$

\item $p(v_{i})$ $L$ $d_{1}$ $L$ $p(w)$ $L$ $p(u_{j})$ $L$ $d_{2}.$
\end{itemize}

As $R=129\left\vert \mathcal{C}\right\vert +464\left\vert \mathcal{D}%
\right\vert $, it follows that for every $\left( u_{i},v_{j},u_{k}\right)
\in \mathcal{C}$, either $u_{i}$ $L^{\prime }$ $v_{j}$ $L^{\prime }$ $u_{k}$
or $u_{k}$ $L^{\prime }$ $v_{j}$ $L^{\prime }$ $u_{i}$ holds. Moreover, for
every $\left( v_{i},w,u_{j}\right) \in \mathcal{D}$, we have either $v_{i}$ $%
L^{\prime }$ $w$ $L^{\prime }$ $u_{j}$ or $u_{j}$ $L^{\prime }$ $w$ $%
L^{\prime }$ $v_{i}$. This implies that $L^{\prime }$ is a solution of $I$.
\textbf{\rule{2mm}{2mm}}


\newpage
\section*{Appendix}

The following figures and tables are utilized in the proof of Theorem 7.
In all tables shown below, $(\{a,b\}, c, d)$ represents the lists $(a, b, c, d)$ and $(b, a, c, d)$.


\begin{figure}[h!]
\begin{center}
\includegraphics[width=\textwidth]{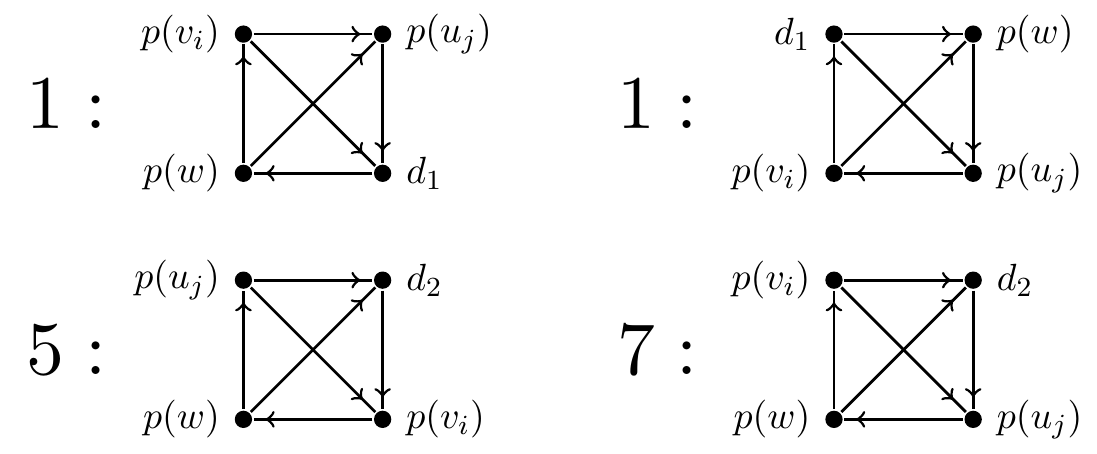}
\end{center}
\caption{This figure shows the top favorite products and the preferences of the 14 buyers created for an $(v_i,w,u_j,)\in \mathcal{D}$ in the NP-hardness of the SE-PA-$f^{SC}$ problem in Theorem~7. The number on the left side of each graph is the number of buyers with the top circle and preferences as shown in the graph on the right side. An arc from a product to another product means that the former one is preferred to the latter one.}
\label{fig_NP_top_circle_4_b}
\end{figure}

%
%
%
%

\begin{table}[h!]
\begin{center}
\begin{tabular}{c|c||c|c}\hline
 list                    &  results    & list   & results\\ \hline

$(\{p(u_i),p(u_k)\},  d_1,  p(v_j))$  &   $p(v_j), d_1$   & $(\{p(u_k),d_1\},  p(u_i),  p(v_j))$ & $p(v_j),p(u_i)$   \\

$(\{p(u_i),p(u_k)\},  p(v_j),  d_1)$ &  $d_1, d_1$&$(\{p(u_k),d_1\},  p(v_j),  p(u_i))$ & $p(v_j),p(u_i)$\\

$(\{p(u_i),d_1\},  p(u_k),  p(v_j))$  & $p(v_j), p(v_j)$ & $(\{p(u_k),p(v_j)\},  p(u_i),  d_1)$ & $d_1,p(u_i)$\\
$(\{p(u_i),d_1\},  p(v_j),  p(u_k))$ &  $p(v_j), p(u_k)$ &$(\{p(u_k),p(v_j)\},  d_1,  p(u_i))$ & $p(u_i),p(u_i)$\\
$(\{p(u_i),p(v_j)\},  p(u_k),  d_1)$ & $d_1,d_1$  & $(\{d_1,p(v_j)\},  p(u_i),  p(u_k))$ & $p(u_i), p(u_k)$ \\

$(\{p(u_i),p(v_j)\},  d_1,  p(u_k))$ & $p(u_k), p(u_k)$ &$(\{d_1,p(v_j)\},  p(u_k),  p(u_i))$ & $p(u_i), p(u_i)$ \\ \hline
\end{tabular}
\end{center}
\caption{This table summarizes all lists of products $p(u_i), p(v_j), p(u_k)$ and $d_1$ chosen by the first two buyers corresponding to a 3-tuple $(u_i,v_j,u_k)\in \mathcal{C}$ (see the two graphs above in Figure~\protect\ref{fig_NP_top_circle_4} for the preferences of these two buyers). The results are the products that the two buyers will choose, one for each.
}
\label{tab_NP_hard_aone}
\end{table}

\begin{table}[h!]
\begin{center}
\begin{tabular}{c|c||c|c|c|c}\hline
 list                    &  results  & list   & results \\ \hline

$(\{p(u_k),d_2\},  p(u_i),  p(v_j))$  &   $p(v_j), p(v_j)$  & $(\{d_2,p(u_i)\},  p(u_k),  p(v_j))$ & $p(v_j),p(v_j)$   \\

$(\{p(u_k),d_2\},  p(v_j),  p(u_i))$ &  $p(u_i), p(v_j)$ &$(\{d_2,p(u_i)\},  p(v_j),  p(u_k))$ & $p(v_j),p(u_k)$\\

$(\{p(u_k),p(u_i)\},  d_2,  p(v_j))$  & $p(v_j), p(v_j)$ & $(\{d_2,p(v_j)\},  p(u_k),  p(u_i))$ & $p(u_i),p(u_i)$\\
$(\{p(u_k),p(u_i)\},  p(v_j),  d_2)$ &  $p(v_j), p(v_j)$ &$(\{d_2,p(v_j)\},  p(u_i),  p(u_k))$ & $p(u_k),p(u_k)$\\
$(\{p(u_k),p(v_j)\},  d_2,  p(u_i))$ & $p(u_i),p(u_i)$  & $(\{p(u_i),p(v_j)\},  p(u_k),  d_2)$ & $p(u_k), d_2$ \\

$(\{p(u_k),p(v_j)\},  p(u_i),  d_2)$ & $d_2, p(u_i)$ &$(\{p(u_i),p(v_j)\},  d_2,  p(u_k))$ & $p(u_k), p(u_k)$ \\ \hline
\end{tabular}
\end{center}
\caption{This table summarizes all lists of products $p(u_i), p(v_j), p(u_k)$ and $d_2$ chosen by the last two buyers corresponding to a 3-tuple $(u_i,v_j,u_k)\in \mathcal{C}$ (see the two graphs below in Figure~\protect\ref{fig_NP_top_circle_4} for the preferences of these two buyers). The results are the products that the two buyers will choose, one for each.
}
\label{tab_NP_hard_atwo}
\end{table}


\begin{table}[h!]
\begin{center}
\begin{tabular}{c|c||c|c}\hline
 list                    &  results    & list   & results\\ \hline

$(\{p(v_i),p(u_j)\},  d_1,  p(w))$  &   $p(w), d_1$   & $(\{p(u_j),d_1\},  p(v_i),  p(w))$ & $p(w),p(v_i)$   \\

$(\{p(v_i),p(u_j)\},  p(w),  d_1)$ &  $d_1, d_1$&$(\{p(u_j),d_1\},  p(w),  p(v_i))$ & $p(w),p(v_i)$\\

$(\{p(v_i),d_1\},  p(u_j),  p(w))$  & $p(w), p(w)$ & $(\{p(u_j),p(w)\},  p(v_i),  d_1)$ & $d_1,p(v_i)$\\
$(\{p(v_i),d_1\},  p(w),  p(u_j))$ &  $p(w), p(u_j)$ &$(\{p(u_j),p(w)\},  d_1,  p(v_i))$ & $p(v_i),p(v_i)$\\
$(\{p(v_i),p(w)\},  p(u_j),  d_1)$ & $d_1,d_1$  & $(\{d_1,p(w)\},  p(v_i),  p(u_j))$ & $p(v_i), p(u_j)$ \\

$(\{p(v_i),p(w)\},  d_1,  p(u_j))$ & $p(u_j), p(u_j)$ &$(\{d_1,p(w)\},  p(u_j),  p(v_i))$ & $p(v_i), p(v_i)$ \\ \hline
\end{tabular}
\end{center}
\caption{This table summarizes all lists of products $p(v_i), p(u_j), p(w)$ and $d_1$ chosen by the first two buyers corresponding to a 3-tuple $(v_i,w,u_j)\in \mathcal{D}$ (see the two graphs above in Figure~\protect\ref{fig_NP_top_circle_4_b} for the preferences of these two buyers). The results are the products that the two buyers will choose.
}
\label{tab_NP_hard_one}
\end{table}

\begin{table}[h!]
\begin{center}
\begin{tabular}{c|c||c|c|c|c}\hline
 list                    &  results  & list   & results \\ \hline

$(\{p(u_j),d_2\},  p(v_i),  p(w))$  &   $p(w), p(w)$  & $(\{d_2,p(v_i)\},  p(u_j),  p(w))$ & $p(w),p(w)$   \\

$(\{p(u_j),d_2\},  p(w),  p(v_i))$ &  $p(v_i), p(w)$ &$(\{d_2,p(v_i)\},  p(w),  p(u_j))$ & $p(w),p(u_j)$\\

$(\{p(u_j),p(v_i)\},  d_2,  p(w))$  & $p(w), p(w)$ & $(\{d_2,p(w)\},  p(u_j),  p(v_i))$ & $p(v_i),p(v_i)$\\
$(\{p(u_j),p(v_i)\},  p(w),  d_2)$ &  $p(w), p(w)$ &$(\{d_2,p(w)\},  p(v_i),  p(u_j))$ & $p(u_j),p(u_j)$\\
$(\{p(u_j),p(w)\},  d_2,  p(v_i))$ & $p(v_i),p(v_i)$  & $(\{p(v_i),p(w)\},  p(u_j),  d_2)$ & $p(u_j), d_2$ \\

$(\{p(u_j),p(w)\},  p(v_i),  d_2)$ & $d_2, p(v_i)$ &$(\{p(v_i),p(w)\},  d_2,  p(u_j))$ & $p(u_j), p(u_j)$ \\ \hline
\end{tabular}
\end{center}
\caption{This table summarizes all lists of products $p(v_i), p(u_j), p(w)$ and $d_2$ chosen by the last 5+7=12 buyers corresponding to a 3-tuple $(v_i,w,u_j)\in \mathcal{D}$ (see the two graphs below in Figure~\protect\ref{fig_NP_top_circle_4_b} for the preferences of these 12 buyers). The results $p,p'$ shown in the table means that first 5 buyers choose $p$ and the last $7$ buyers choose $p'$. So, for a list with results $p,p'$, the total profits of the products chosen by the 12 buyers is $5\mu(p)+7\mu(p')$.
}
\label{tab_NP_hard_two}
\end{table}

\end{document}